\documentclass[aps,pre,twocolumn,groupedaddress,floatfix,showpacs]{revtex4-1}

\usepackage{graphicx}
\usepackage{amsmath}
\usepackage{amssymb}

\providecommand{\boldsymbol}[1]{\mbox{\boldmath $#1$}}

\begin{document}

\title{Event Driven Langevin simulations of Hard Spheres}

\author{A. Scala}
\affiliation{ISC-CNR Dipartimento di Fisica, Sapienza Universit\`a di Roma
Piazzale Moro 5, 00185 Roma, Italy}
\affiliation{London Institute of Mathematical Sciences, 22 South Audley St
Mayfair London W1K 2NY, UK}

\date{\today}

\begin{abstract}
The blossoming of interest in colloids and nano-particles has given
renewed impulse to the study of hard-body systems. In particular,
hard spheres have become a real test system for theories and experiments.
It is therefore necessary to study the complex dynamics of such systems
in presence of a solvent; disregarding hydrodynamic interactions,
the simplest model is the Langevin equation. Unfortunately,
standard algorithms for the numerical integration of the Langevin
equation require that interactions are slowly varying during an
integration time-step. This in not the case for hard-body systems,
where there is no clearcut between the correlation time of the noise
and the time-scale of the interactions. Starting first from a splitting
of the Fokker-Plank operator associated with the Langevin dynamics,
and then from an approximation of the two-body Green's function, we
introduce and test two new algorithms for the simulation of the Langevin
dynamics of hard-spheres.
\end{abstract}

\pacs{05.10.-a,05.40.Jc,05.10.Gg}

\maketitle

\section{Introduction}

Hard spheres (HS) are a reference system for structural and dynamical
theories of fluids \cite{Andersen71,HansenMcDonaldBook}, but idealized:
the infinitely steep potential is essentially a way of capturing the
effects of steric interactions. On the atomic or the molecular scale
two body interactions are mostly modelled by Lennard-Jones or Coulumb
potentials; experiments on colloids shift the length scales of interest
up to roughly $1\, nm$ to $1000\, nm$ where objects can behave as
hard bodies and are still small enough to exhibit thermal or Brownian
motion in a solvent. Dynamical light scattering \cite{BerneBook,DeGiorgioBook}
has already provided a rich collection of data for such systems, encouraging
a considerable effort in understanding the dynamics; the possibility
of following single particle trajectories via confocal microscopy
of latex particle \cite{vanBlaaderen95} has allowed a direct view
on an experimental realization of HS systems and their dynamics \cite{Kegel00,Weeks00}.

The simplest model of a suspension of neutral particles is to consider
a system of HS in an ideal solvent with no hydrodynamic interactions;
real suspensions are often described in terms of their deviations
from such ideal system. This is the most interesting model for theoreticians
and many results have been derived: the two body case (and hence the
low density case) has been solved exactly \cite{Hanna82,Ackerson82},
while at moderate and high packing fractions various Enskog-like 
\cite{Felderhof83A,Cichocki90theo}
or Mode Coupling theories \cite{Hess83,FuchsPRL2002} have been applied
to understand the dynamics. 
While hydrodynamic interactions (HI) are
well understood at low particle densities, much less is known at high
densities, and theories often proceed by disregarding them \cite{Felderhof83B}. 
As an example,
theories regarding glass transition often disregard HI effect, like in the case 
of the Mode Coupling theory for Brownian hard spheres\cite{FuchsPRL2002,BraderPRL2007,BraderPRL2008} 
or Brownian hard discs in shear flow\cite{Henrich2009}. 

Non-HI simulations therefore have their place in testing such theories,
and in circumventing the huge effort needed to simulate HI 
\cite{Brady88,Rjoy05,Groot97,Tanaka00}.

In order to validate non-HI theories for HSs it is necessary to use computer
simulations, as only a qualitative agreement is to be expected among
non-HI theories and data for real suspension. Standard simulation
methods for Brownian dynamic like the well-known Ermak-McCammon \cite{Ermak78}
require continuous potentials; to circumvent such problem several
algorithms have been introduced with various degrees of justification
\cite{Cichocki90sim,Heyes93,Sillescu94,Foss2000} for the over-damped
dynamics; only recently it has been recognized that in the case of
hard interactions such simulations are better performed by event-driven
(ED) codes \cite{Strating99,Tao06,BD4HS}. We introduce two new ED
algorithms that go beyond the over-damped approximation and allow 
for the simulation of the full Brownian dynamics of HSs.

\section{Methods}

We consider a system of $N$ HSs governed by the Langevin equation 
\begin{equation}
\left\{ \begin{array}{ccc}
\partial_{t}\mathbf{v}_{i} & = & -\gamma\mathbf{v}_{i}+\boldsymbol{\mathbf{a}_{i}}+\boldsymbol{\xi_{i}}\\
\partial_{t}\boldsymbol{r}_{i} & = & \mathbf{v}_{i}\end{array}\right.\label{eq:LangManyBody}\end{equation}
for the positions $\mathbf{r}_{i}$ and the velocities $\mathbf{v}_{i}$; 
here $\gamma$ is the friction constant, $\mathbf{a}_{i}=-m^{-1}\partial_{\mathbf{x}}U$
the acceleration, $m$ is the mass of the HSs, $U$ is the potential
energy and $m\boldsymbol{\xi}_{i}$ are the zero-mean random forces
due to the solvent. We assume that such random forces are delta correlated
and satisfy the fluctuation-dissipation theorem 
\begin{equation}
\left\langle \boldsymbol{\xi}_{i}(\mathbf{x},t)\otimes\boldsymbol{\xi}_{j}(\mathbf{x'},t')\right\rangle =
\gamma\frac{2k_{B}T}{m}\delta\left(\mathbf{x}-\mathbf{x'}\right)\delta\left(t-t'\right)\delta_{ij}\boldsymbol{1}
\end{equation}

In the case of continuous interactions, it is possible to define
stochastic Taylor expansions \cite{KannanBook}; correspondingly, integration schemes
of the $k$-th order with errors of order $(\Delta t)^{k}$ in the
time-step$\Delta t$ can be introduced \cite{KloedenPlatenBook}.
In the case of hard-body interactions, all the standard machinery of
stochastic calculus breaks down due to the singular nature of the
interaction potential and new methods must be developed.

We consider the Fokker-Plank equation associated to the SDE (\ref{eq:LangManyBody})
(Kramers' equation \cite{Kramers1940})
\begin{equation}
\partial_{t}W = \mathbf{L}_{K} W
\label{eq:Kramers}
\end{equation}
where $W\left(\mathbf{r},\mathbf{v},t\right)$ is the probability distribution function (PDF) for the
positions $\mathbf{r}=\left\{ \mathbf{r}_{i}\right\} $ 
and the velocities $\mathbf{v}=\left\{ \mathbf{v}_{i}\right\} $ of the particles, 
$v_{th}^{2}=k_{B}T/m$ relates to the temperature and 
\begin{equation}
\mathbf{L}_{K} =
\gamma\left(
	\partial_{\mathbf{v}}\cdot\mathbf{v} +
		v_{th}^{2}\partial_{\mathbf{v}}^{2}
\right) - \left(
	\mathbf{v}\cdot\partial_{\mathbf{r}} +
	\mathbf{a}\cdot\partial_{\mathbf{v}}
\right)
\label{eq:KramerOP}
\end{equation}
is the Kramer operator. Integrating the SDE (\ref{eq:LangManyBody})
for a finite time-step $\Delta t$ corresponds to extracting a configuration
$\left\{ \mathbf{r}^{t+\Delta t},\mathbf{v}^{t+\Delta t}\right\} $
according to the probability $e^{\mathbf{L}_{K}\Delta t}\delta\left(\mathbf{x}-\mathbf{x}^{t},\mathbf{v}-\mathbf{v}^{t}\right)$.

\section{Splitted Brownian Dynamics}

To obtain a numerical approximation, a powerful approach is to split
the evolution operator $e^{\mathbf{L}_{K}\Delta t}$
in a product $e^{\mathbf{L}_{K}\Delta t}\approx{\displaystyle \prod_{i}}e^{a_{i}\mathbf{L}_{i}\Delta t}$
of exactly-integrable operators  $\mathbf{L}_{i}$
\cite{Forbert00} ensuring that that the decomposition is positive
(i.e. all $a_{i}>0$) \cite{Chin05}. Therefore, to each splitting corresponds an algorithm
in which in a single time-step $\Delta t$ the operators $e^{a_{i}\mathbf{L}_{i}\Delta t}$
are applied in sequence.

We first choose to split $\mathbf{L}_{K}$ into the reversible (or streaming) operator 
$\mathbf{L}_{rev}=-\left(\mathbf{v}\cdot\partial_{\mathbf{r}}+\mathbf{a}\cdot\partial_{\mathbf{v}}\right)$
and the irreversible (or collision) operator $\mathbf{L}_{irr}=\gamma\left(\partial_{\mathbf{v}}\cdot\mathbf{v}+v_{th}^{2}\partial_{\mathbf{v}}^{2}\right)$
\cite{RiskenBook}; we indicate the corresponding algorithm as Splitted Brownian Dynamics (SBD). 

The operator $\mathbf{L}_{rev}$ is the Liouvillian associated to
the Hamiltonian $\mathcal{H}=m\mathbf{\, v}\cdot\mathbf{v}/2+U$.
In the case of step potentials, the associated reversible equation
of motion can be integrated via event-driven molecular dynamics (EDMD)
\cite{RapaBook} with a precision limited only by the numerical round-off
errors; therefore the propagator $e^{\mathbf{L}_{rev}\Delta t}$ can
be implemented with extreme accuracy.

The operator $\mathbf{L}_{irr}$ corresponds to the interaction with the bath; 
the associated SDE is $\partial_{t}\mathbf{v}=-\gamma\mathbf{v}+\xi$
can be exactly integrated giving an explicit formula for 
the evolution $\mathbf{v}^{t+\Delta t}=e^{\mathbf{L}_{irr}\Delta t}\mathbf{v}^{t}$:
\begin{equation}
\mathbf{v}_{i,\alpha}^{t+\Delta t}=e^{-\gamma\Delta t}\mathbf{v}_{i,\alpha}^{t}+\sqrt{v_{th}^{2}\left(1-e^{-2\gamma\Delta t}\right)}\Gamma
\label{eq:ThermalizeVel}
\end{equation}
where $\Gamma$ is a unitary Gaussian random variable and $\alpha\in\left\{ x,y,z\right\} $. 

The algorithm for the single SBD time-step 
$e^{\mathbf{L}_{rev}\Delta t}e^{\mathbf{L}_{irr}\Delta t}$
consists therefore in an EDMD simulation \cite{RapaBook} of length
$\Delta t$ followed by a thermalization of the velocities according
to eq.(\ref{eq:ThermalizeVel}). We notice that the error is at most
quadratic (as can be checked via Taylor expansion $e^{\mathbf{L}_{rev}\Delta t}e^{\mathbf{L}_{irr}\Delta t}=e^{\mathbf{L}_{K}\Delta t}+\mathcal{O}\left(\Delta t^{2}\right)$
) \emph{and regards only in the dynamics}; in fact, SBD is equivalent
(upon identifying the angle $\alpha$ mixing reversible and
irreversible evolution with $\cos\left(\alpha\right)=e^{-\gamma(t-t')}$)
to the Generalized Hybrid Monte Carlo \cite{Kennedy99} and therefore
explores the canonical ensemble as long as the propagation steps 
$e^{\mathbf{L}_{rev}\Delta t}$,$e^{\mathbf{L}_{irr}\Delta t}$
can be exactly implemented (as in our case).

\begin{figure}
\includegraphics[width=0.8\linewidth,keepaspectratio]{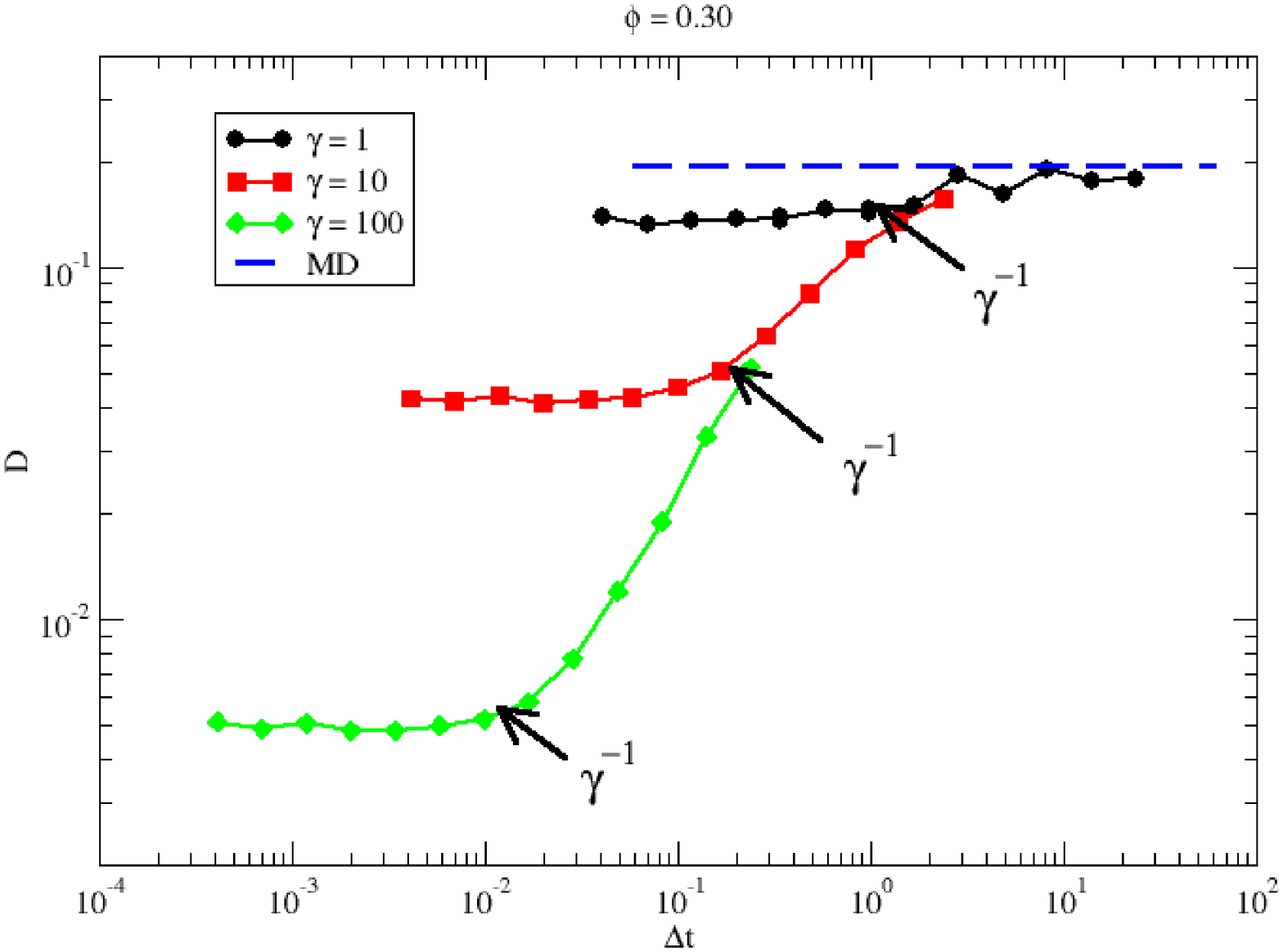}
\caption{\label{fig:Effect-of-eta}
Effect of the damping coefficient $\gamma$
on the size of the simulation step $\Delta t$ (all quantities in reduced units). 
The diffusion coefficient
$D$ from simulations is plotted versus the time-step size $\Delta t$
for various $\gamma$'s. As expected, the system approaches the $MD$
value for diffusion regardless of $\gamma$ for $\Delta t\rightarrow\infty$.
The ``true'' value of $D$ is obtained for $\Delta t\to0$ . 
We observe at small $\Delta t$'s a plateau in the $D$~vs~$\Delta t$
plot for $\Delta t\lesssim\gamma^{-1}$, signalling that the ``true''
value of $D$ is approached. Results are presented for packing fraction $\phi=0.30$;
a completely analogous behaviour is found
at a low packing fraction $\phi=0.10$ and an high packing fraction
$\phi=0.45$.
}

\end{figure}

It is therefore of interest to give some physical bounds on the magnitude
of the feasible time-step $\Delta t$. First, we notice that for $\Delta t\to\infty$
the dynamics reduces to and MD simulations where velocities are extracted
each $\Delta t$ from a Maxwellian; therefore if the time-step is much
bigger than the average inter-particle collision time, results of
classical MD are to be expected. Accordingly, we find that for big $\Delta t$ 
the algorithm overestimates the diffusion coefficient (fig. \ref{fig:Effect-of-eta} ); 
this is to be expected as the mean free path (in absence of collisions) 
of a particle is of order
$v_{th}\Delta t$ instead of $\gamma^{-1}v_{th}\sqrt{\Delta t}$.
Second, the magnitude of $\Delta t$ is naturally bounded the damping
time $\tau=\gamma^{-1}$ ; therefore the SBD is not well indicated
for simulations in the over-damped limit $\gamma/m\to\infty$. 
Accordingly, we find that SBD overestimates diffusion coefficients 
for $\Delta t \gtrsim \gamma^{-1}$ (fig.\ref{fig:Effect-of-eta}); 
it is therefore necessary to develop an alternative approach 
for the simulation of systems with high damping.

\section{Approximate Green's function dynamics}

It has been shown in \cite{BD4HS} that the over-damped limit of eq.(\ref{eq:LangManyBody})
can be simulated efficiently using ED codes\cite{BD4HS}.
The algorithm relies on considering time steps $\Delta t$ small enough so that mostly
binary collisions are relevant, i.e. the average displacement should be less than
the average inter-particle separation. Moreover, average displacement should be smaller than 
the HSs' radii in order to map the interaction of two nearby HSs in the problem of a random walk 
near a reflective wall. Under such approximations, the 
true two-body stochastic dynamics for over-damped Brownian HSs 
can be implemented by algorithm of \cite{BD4HS} 
in which each step consists in predicting the displacements $\Delta\mathbf{x}$
of the HSs via the free propagator, introducing fictive
velocities $\mathbf{v}^{f}=\Delta\mathbf{x}/\Delta t$, and performing
an EDMD with such fictive velocities during $t$ and $t+\Delta t$. 
We extend such approach to the general Brownian case.

First, we need to predict the positions of the HSs after a time-step $\Delta t$
according to their free propagation, i.e. the solution of eq.(\ref{eq:LangManyBody})
with no interaction ($\mathbf{a}=0)$:
\begin{equation}
\left\{ \begin{array}{ccc}
\mathbf{v}^{t+\Delta t} & = & \mathbf{v}^{t}+\overline{\Delta\mathbf{v}}+\Delta\mathbf{v}_{R}\\
\mathbf{r}^{t+\Delta t} & = & \mathbf{r}^{t}+\overline{\Delta\mathbf{r}}+\Delta\mathbf{r}_{R}\end{array}\right.\label{eq:1BGF}\end{equation}
The particle displacements contain both systematic parts  
$\overline{\Delta\mathbf{v}}=(e^{-\gamma t}-1)\,\mathbf{v}^{t}$,
$\overline{\Delta\mathbf{r}}=\gamma^{-1}(1-e^{-\gamma t})\,\mathbf{v}^{t}$
and stochastic displacements. The stochastic displacements
$\Delta\mathbf{v}_{R}$, $\Delta\mathbf{r}_{R}$ 
are zero-mean correlated gaussian variables with variances 
$\left\langle \Delta\mathbf{v}_{R}^{2}\right\rangle =m^{-1}k_{B}T\,(1-e^{-2\gamma t})$,
$\left\langle \Delta\mathbf{r}_{R}^{2}\right\rangle =\gamma^{-1}m^{-1}k_{B}T\,\left[2t-\gamma^{-1}\left(3+4e^{-\gamma t}+e^{-2\gamma t}\right)\right]$
and cross-correlation 
$\left\langle \Delta\mathbf{r}_{R}\Delta\mathbf{v}_{R}\right\rangle =\gamma^{-1}m^{-1}k_{B}T\,\left(1-e^{-\gamma t}\right)^{2}$
\cite{ATBook}.

\begin{figure}
\includegraphics[width=0.8\linewidth,keepaspectratio]{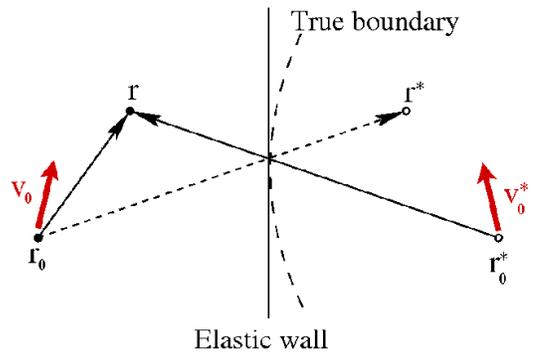}
\caption{\label{fig:Image-method}
A two body problem for hard spheres can be mapped into the problem of a point particle 
interacting with a larger sphere. When particles are very near, the problem further simplifies 
to the interaction of a Langevin particle with a reflective flat wall, whose solution can be  
derived by applying the Image Method to the Langevin equation. 
In fact, the Green function must zero inside the
wall and must satisfy the no-flux boundary conditions at the wall. Combining
the free Green function of the particle in its initial position 
and the free Green function of its image 
(with the normal-to-the-wall component of the initial velocity reflected) 
satisfies both Kramer' equations and reflective boundary conditions giving the correct solution.}
\end{figure}

If we consider a time-step such that the average displacement is less than
the average inter-particle separation, we can consider only the corrections due to two-body interactions. 
In the limit of small $\Delta t$, a couple of HSs will interact only when they
start from nearby positions. In particular, if $\gamma^{-1}v_{th}\sqrt{\Delta t}\ll\sigma$,
i.e. the average free displacement is much smaller than the diameter
$\sigma$ of the HSs, the dynamics of two particles $A$ and $B$
can be approximated as the Langevin dynamics of a point particle at
a distance $\left(\mathbf{r}_{A}-\mathbf{r}_{B}\right)\left(1-\sigma/\left\Vert \mathbf{r}_{A}-\mathbf{r}_{B}\right\Vert \right)$
from a flat wall. It is possible to solve such problem with a straightforward generalization of the image
method applied in \cite{BD4HS}. In fact, the solution given by the free particle Green's function plus an image
particle with a reflected velocity beyond the reflective wall (fig. \ref{fig:Image-method})
correctly satisfies the zero-current boundary condition $\left.\hat{\mathbf{n}} \cdot \mathbf{j}\right|_{wall}=0$, where $\hat{\mathbf{n}}$ is the normal to the wall and $\mathbf{j}(\mathbf{r},t)=\int  \mathbf{v} W(\mathbf{r},\mathbf{v},t) d\mathbf{v}$ is the probability current for the position.

Such solution can be implemented exactly by predicting the new positions and velocities 
$\mathbf{r}^{t+\Delta t}$,$\mathbf{v}^{t+\Delta t}$ according to eq.(\ref{eq:1BGF}), 
defining fictive velocities $\mathbf{v}^{f}=\left(\mathbf{r}^{t+\Delta t}-\mathbf{r}^{t}\right)/\Delta t$
and performing an EDMD simulation with such fictive velocities during
$\Delta t$; if a collision happens, the component of the relative
velocity normal to the contact point must be reflected for both the
fictive $\mathbf{v}^{f}$ and the predicted velocities $\mathbf{v}^{t+\Delta t}$. 
We indicate such algorithm as the approximate Green's function dynamics
(AGD). In the over-damped limit, the prediction of the velocities and
positions decorrelates and the algorithm correctly reduces to the over-damped
case of \cite{BD4HS}.

As for the SBD algorithm, it can be proven that the AGD scheme respects
detailed balance and ergodicity and therefore explores the correct
ensemble for HSs; hence, errors are again only in dynamic quantities.
At difference with SBD, we have no analytic estimate for the error;
nevertheless, we expect that the the mean-free path in absence of collisions
$\gamma^{-1}v_{th}\sqrt{\Delta t}$ must be smaller than the
radius of the HSs in order to satisfy the flat-wall approximation,
and must be smaller than the average inter-particle distance
in order to avoid multiple collisions (hence higher than two-body
effects) during $\Delta t$.

\begin{figure}
\includegraphics[width=0.7\linewidth,keepaspectratio]{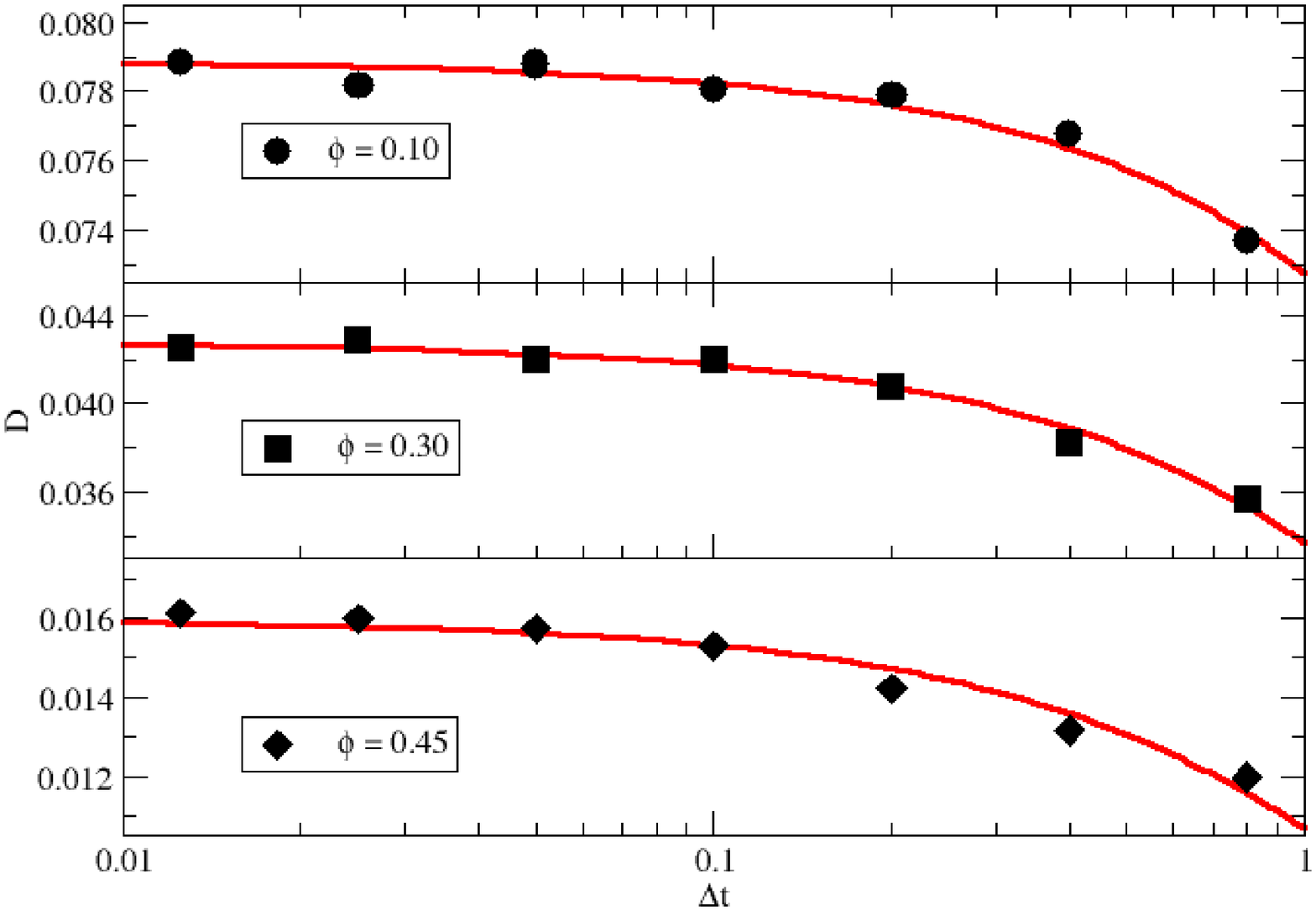}
\includegraphics[width=0.7\linewidth]{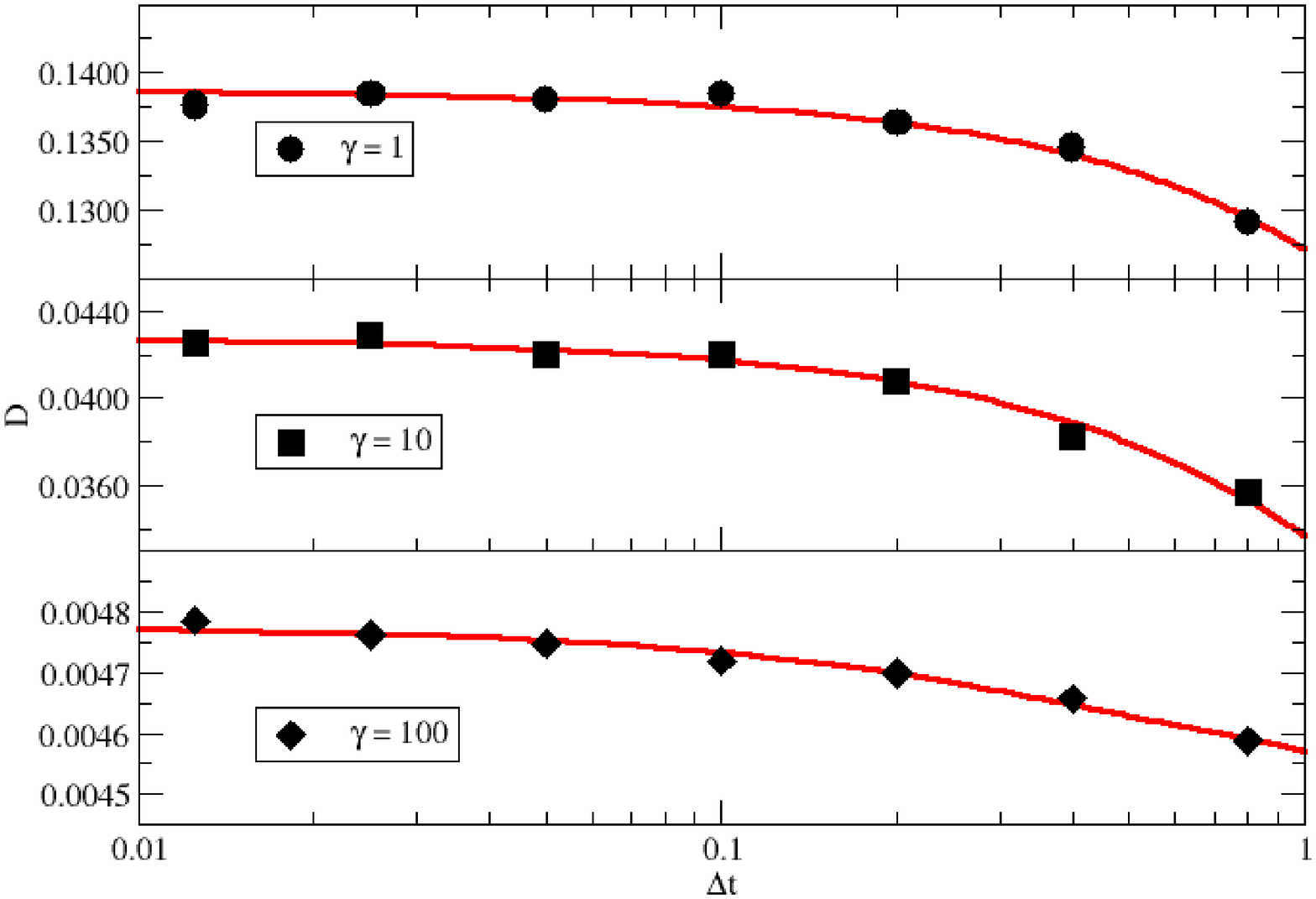}
\caption{\label{fig:GreenDtBehavior}
Effects of packing fraction $\phi$ (upper panel) and of the damping coefficient 
$\gamma$ (lower panel) on the time-step $\Delta t$ for
the AGF algorithm. 
All quantities in reduced units; 
thick lines are just a guide for the eye. 
Diffusion $D$ is calculated averaging over $10$
independent trajectories for $2000$ particle systems; 
simulations are long at least $10$ times the structural correlation time.
In the upper panel, results are shown for $\phi=0.10,0.30, 0.45$ at fixed damping $\gamma=10$.
In the lower panel, results are shown for $\gamma=1,10,100$ 
at fixed packing fraction $\phi=0.30$.
Notice that the estimated diffusion coefficient $D$ has a small relative variation 
in the wide range of dampings $\gamma$s and packing fractions $\phi$s analysed. 
As a rule of thumb, to estimate $D$ with an accuracy much smaller than $1\%$ 
time-step of order $\Delta t \sim 0.1$ are already enough.}
\end{figure}

In order to check that the behaviour of AGS is driven just by geometrical
considerations, we have simulated HS systems at different $\gamma$ and $\phi$
varying the time-step $\Delta t$ in the range $\left[10^-2,10^0\right]$ (reduced units). 
At difference with SBD where diffusion can vary even by a order of magnitude in such a $\Delta t$ range, 
the values of $D$ measured from AGD vary a few percent over the range 
and long simulations are been necessary to have enough statistics to detect
the behaviour of $D$ that would otherwise look flat. 
In fig.~\ref{fig:GreenDtBehavior} we show that the measured
diffusion coefficient $D$ versus the AGS simulation time-step
displays a plateau (i.e. fluctuations become much smaller than $1\%$) 
already for $\Delta t\lesssim 0.1$ regardless of $\gamma$ and $\phi$.

\section{Conclusions}

Hard spheres, and in general hard body systems in suspension, have become a realistic model due to the
developments of experimental techniques for the investigation of colloidal systems and nano-particles; 
yet the dynamics of such systems is hard to simulate via the standard Brownian dynamics algorithms. 
In fact, classical continuous-time algorithms fail due to instantaneous character of the interactions; 
we have shown instead how it is possible to simulate the full Langevin dynamics of Hard Spheres.

First, we have shown how the simplest splitting of the stochastic evolution operator 
(a technique often referred to as "Trotterization" from Trotter's seminal work\cite{TrotterPAMS59}) 
allows to write an algorithm (the Splitted Brownian Dynamics - SBD).
The SBD algorithm  becomes inefficient of high viscosities but via the operator-splitting technique
could easily take account for the interaction with external fields or 
with the presence of fluxes (like shear) in the surrounding fluid.

Second, we have shown how by considering the two body dynamics of Brownian Hard Spheres 
it is possible  do develop an algorithm (the Approximate Green's function Dynamics AGD) 
that overcomes such problem and works equally well for a wide range 
of packing fractions and viscosities.
To develop the AGD algorithm, we have solved the problem of the Langevin dynamics 
$\partial_{t}v  =  -\gamma v + \xi$ of a point particle in presence of a reflective wall 
by extending the classical Image Method solution for the over-damped Brownian dynamics 
$\partial_{t}x  =  \eta$ of a point particle in presence of a reflective wall
(here $\xi$, $\eta$ are noises).
The AGD algorithm is Event Driven and considers fictive collisions between Hard Spheres.
While it should possible to take into account the polydispersity of a system
by considering also effective masses in the fictive collisions as in \cite{BD4HS},
including shear or external fields in the AGD algorithm looks more complicated 
as it would require the solution of the particle - reflective wall problem with 
external fields/shear.

Both SBD and AGD simulations explore the canonical ensemble for Hard Spheres and therefore reproduce 
the correct equilibrium thermodynamics. 
They belong to the class of Asynchronous Event-Driven Particle Algorithms\cite{DonevSIM2009} and can 
be easily implemented by adapting existing codes for ED dynamics \cite{RapaBook} or 
Brownian Dynamics \cite{HardBrown} of Hard Spheres.

\begin{acknowledgments}
The author thanks Th. Voigtmann for long and useful discussions.
\end{acknowledgments}

\bibliography{LANG4HS}

\begin{thebibliography}{10}%
\makeatletter
\providecommand \@ifxundefined [1]{%
 \ifx #1\undefined \expandafter \@firstoftwo
 \else \expandafter \@secondoftwo
\fi
}%
\providecommand \@ifnum [1]{%
 \ifnum #1\expandafter \@firstoftwo
 \else \expandafter \@secondoftwo
\fi
}%
\providecommand \enquote [1]{``#1''}%
\providecommand \bibnamefont  [1]{#1}%
\providecommand \bibfnamefont [1]{#1}%
\providecommand \citenamefont [1]{#1}%
\providecommand\href[0]{\@sanitize\@href}%
\providecommand\@href[1]{\endgroup\@@startlink{#1}\endgroup\@@href}%
\providecommand\@@href[1]{#1\@@endlink}%
\providecommand \@sanitize [0]{\begingroup\catcode`\&12\catcode`\#12\relax}%
\@ifxundefined \pdfoutput {\@firstoftwo}{%
 \@ifnum{\z@=\pdfoutput}{\@firstoftwo}{\@secondoftwo}%
}{%
 \providecommand\@@startlink[1]{\leavevmode\special{html:<a href="#1">}}%
 \providecommand\@@endlink[0]{\special{html:</a>}}%
}{%
 \providecommand\@@startlink[1]{%
  \leavevmode
  \pdfstartlink
   attr{/Border[0 0 1 ]/H/I/C[0 1 1]}%
   user{/Subtype/Link/A<</Type/Action/S/URI/URI(#1)>>}%
  \relax
 }%
 \providecommand\@@endlink[0]{\pdfendlink}%
}%
\providecommand \url  [0]{\begingroup\@sanitize \@url }%
\providecommand \@url [1]{\endgroup\@href {#1}{\urlprefix}}%
\providecommand \urlprefix [0]{URL }%
\providecommand \Eprint[0]{\href }%
\@ifxundefined \urlstyle {%
  \providecommand \doi [1]{doi:\discretionary{}{}{}#1}%
}{%
  \providecommand \doi [0]{doi:\discretionary{}{}{}\begingroup
  \urlstyle{rm}\Url }%
}%
\providecommand \doibase [0]{http://dx.doi.org/}%
\providecommand \Doi[1]{\href{\doibase#1}}%
\providecommand \bibAnnote [3]{%
  \BibitemShut{#1}%
  \begin{quotation}\noindent
    \textsc{Key:}\ #2\\\textsc{Annotation:}\ #3%
  \end{quotation}%
}%
\providecommand \bibAnnoteFile [2]{%
  \IfFileExists{#2}{\bibAnnote {#1} {#2} {\input{#2}}}{}%
}%
\providecommand \typeout [0]{\immediate \write \m@ne }%
\providecommand \selectlanguage [0]{\@gobble}%
\providecommand \bibinfo [0]{\@secondoftwo}%
\providecommand \bibfield [0]{\@secondoftwo}%
\providecommand \translation [1]{[#1]}%
\providecommand \BibitemOpen[0]{}%
\providecommand \bibitemStop [0]{}%
\providecommand \bibitemNoStop [0]{.\EOS\space}%
\providecommand \EOS [0]{\spacefactor3000\relax}%
\providecommand \BibitemShut [1]{\csname bibitem#1\endcsname}%
\bibitem{Andersen71}%
  \BibitemOpen
  \bibfield{author}{%
  \bibinfo {author} {\bibfnamefont{H.~C.}\ \bibnamefont{Andersen}}, \bibinfo
  {author} {\bibfnamefont{J.~D.}\ \bibnamefont{Weeks}},\ and\ \bibinfo {author}
  {\bibfnamefont{D.}~\bibnamefont{Chandler}},\ }%
  \bibfield{journal}{%
  \bibinfo {journal} {Physical Review A}\ }%
  \textbf{\bibinfo {volume} {4}},\ \bibinfo {pages} {1597} (\bibinfo {month}
  {Oct.}\ \bibinfo {year} {1971})%
  \bibAnnoteFile{NoStop}{Andersen71}%
\bibitem{HansenMcDonaldBook}%
  \BibitemOpen
  \bibfield{author}{%
  \bibinfo {author} {\bibfnamefont{J.~P.}\ \bibnamefont{Hansen}}\ and\ \bibinfo
  {author} {\bibfnamefont{I.~R.}\ \bibnamefont{McDonald}},\ }%
  \emph{\bibinfo {title} {Theory of Simple Liquid}},\ \bibinfo {edition} {2nd}\
  ed.\ (\bibinfo {publisher} {Academic Press},\ \bibinfo {address} {New York},\
  \bibinfo {year} {1989})%
  \bibAnnoteFile{NoStop}{HansenMcDonaldBook}%
\bibitem{BerneBook}%
  \BibitemOpen
  \bibfield{author}{%
  \bibinfo {author} {\bibfnamefont{B.~J.}\ \bibnamefont{Berne}}\ and\ \bibinfo
  {author} {\bibfnamefont{R.}~\bibnamefont{Pecora}},\ }%
  \emph{\bibinfo {title} {Dynamic Light Scattering: with Applications to
  Chemistry Biology, and Physics}}\ (\bibinfo {publisher} {John Wiley \& Sons,
  New York},\ \bibinfo {year} {1976})%
  \bibAnnoteFile{NoStop}{BerneBook}%
\bibitem{DeGiorgioBook}%
  \BibitemOpen
  \bibfield{author}{%
  \bibinfo {author} {\bibfnamefont{V.}~\bibnamefont{De~Giorgio}}, \bibinfo
  {author} {\bibfnamefont{M.}~\bibnamefont{Corti}},\ and\ \bibinfo {author}
  {\bibfnamefont{M.}~\bibnamefont{Giglio}},\ }%
  \emph{\bibinfo {title} {Light Scattering in Liquids and Macromolecular
  Solutions}}\ (\bibinfo {publisher} {Plenum, New York},\ \bibinfo {year}
  {1980})%
  \bibAnnoteFile{NoStop}{DeGiorgioBook}%
\bibitem{vanBlaaderen95}%
  \BibitemOpen
  \bibfield{author}{%
  \bibinfo {author} {\bibfnamefont{A.}~\bibnamefont{van Blaaderen}}\ and\
  \bibinfo {author} {\bibfnamefont{P.}~\bibnamefont{Wiltzius}},\ }%
  \bibfield{journal}{%
  \bibinfo {journal} {Science}\ }%
  \textbf{\bibinfo {volume} {270}},\ \bibinfo {pages} {1177} (\bibinfo {year}
  {1995})%
  \bibAnnoteFile{NoStop}{vanBlaaderen95}%
\bibitem{Kegel00}%
  \BibitemOpen
  \bibfield{author}{%
  \bibinfo {author} {\bibfnamefont{W.~K.}\ \bibnamefont{Kegel}}\ and\ \bibinfo
  {author} {\bibfnamefont{A.}~\bibnamefont{van Blaaderen}},\ }%
  \bibfield{journal}{%
  \bibinfo {journal} {Science}\ }%
  \textbf{\bibinfo {volume} {287}},\ \bibinfo {pages} {290} (\bibinfo {year}
  {2000})%
  \bibAnnoteFile{NoStop}{Kegel00}%
\bibitem{Weeks00}%
  \BibitemOpen
  \bibfield{author}{%
  \bibinfo {author} {\bibfnamefont{E.~R.}\ \bibnamefont{Weeks}}, \bibinfo
  {author} {\bibfnamefont{J.~C.}\ \bibnamefont{Crocker}}, \bibinfo {author}
  {\bibfnamefont{A.}~\bibnamefont{Levitt}, \bibfnamefont{A.C.and~Schofield}},\
  and\ \bibinfo {author} {\bibfnamefont{D.~A.}\ \bibnamefont{Weitz}},\ }%
  \bibfield{journal}{%
  \bibinfo {journal} {Science}\ }%
  \textbf{\bibinfo {volume} {287}},\ \bibinfo {pages} {627} (\bibinfo {year}
  {2000})%
  \bibAnnoteFile{NoStop}{Weeks00}%
\bibitem{Hanna82}%
  \BibitemOpen
  \bibfield{author}{%
  \bibinfo {author} {\bibfnamefont{S.}~\bibnamefont{Hanna}}, \bibinfo {author}
  {\bibfnamefont{W.}~\bibnamefont{Hess}},\ and\ \bibinfo {author}
  {\bibfnamefont{R.}~\bibnamefont{Klein}},\ }%
  \bibfield{journal}{%
  \bibinfo {journal} {Physica A}\ }%
  \textbf{\bibinfo {volume} {111}},\ \bibinfo {pages} {181} (\bibinfo {month}
  {Mar.}\ \bibinfo {year} {1982})%
  \bibAnnoteFile{NoStop}{Hanna82}%
\bibitem{Ackerson82}%
  \BibitemOpen
  \bibfield{author}{%
  \bibinfo {author} {\bibfnamefont{B.~J.}\ \bibnamefont{Ackerson}}\ and\
  \bibinfo {author} {\bibfnamefont{L.}~\bibnamefont{Fleishman}},\ }%
  \bibfield{journal}{%
  \bibinfo {journal} {Journal of Chemical Physics}\ }%
  \textbf{\bibinfo {volume} {76}},\ \bibinfo {pages} {2675} (\bibinfo {month}
  {Mar.}\ \bibinfo {year} {1982})%
  \bibAnnoteFile{NoStop}{Ackerson82}%
\bibitem{Felderhof83A}%
  \BibitemOpen
  \bibfield{author}{%
  \bibinfo {author} {\bibfnamefont{B.}~\bibnamefont{Felderhof}}\ and\ \bibinfo
  {author} {\bibfnamefont{R.}~\bibnamefont{Jones}},\ }%
  \bibfield{journal}{%
  \bibinfo {journal} {Physica A Statistical Mechanics and its Applications}\ }%
  \textbf{\bibinfo {volume} {121}},\ \bibinfo {pages} {329} (\bibinfo {month}
  {Aug.}\ \bibinfo {year} {1983})%
  \bibAnnoteFile{NoStop}{Felderhof83A}%
\bibitem{Cichocki90theo}%
  \BibitemOpen
  \bibfield{author}{%
  \bibinfo {author} {\bibfnamefont{B.}~\bibnamefont{Cichocki}}\ and\ \bibinfo
  {author} {\bibfnamefont{B.~U.}\ \bibnamefont{Felderhof}},\ }%
  \bibfield{journal}{%
  \bibinfo {journal} {Physical Review A}\ }%
  \textbf{\bibinfo {volume} {42}},\ \bibinfo {pages} {6024} (\bibinfo {month}
  {Nov.}\ \bibinfo {year} {1990})%
  \bibAnnoteFile{NoStop}{Cichocki90theo}%
\bibitem{Hess83}%
  \BibitemOpen
  \bibfield{author}{%
  \bibinfo {author} {\bibfnamefont{W.}~\bibnamefont{Hess}}\ and\ \bibinfo
  {author} {\bibfnamefont{R.}~\bibnamefont{Klein}},\ }%
  \bibfield{journal}{%
  \bibinfo {journal} {Adv.~Phys.}\ }%
  \textbf{\bibinfo {volume} {32}},\ \bibinfo {pages} {173} (\bibinfo {year}
  {1983})%
  \bibAnnoteFile{NoStop}{Hess83}%
\bibitem{FuchsPRL2002}%
  \BibitemOpen
  \bibfield{author}{%
  \bibinfo {author} {\bibfnamefont{M.}~\bibnamefont{Fuchs}}\ and\ \bibinfo
  {author} {\bibfnamefont{M.~E.}\ \bibnamefont{Cates}},\ }%
  \bibfield{journal}{%
  \Doi{10.1103/PhysRevLett.89.248304}{\bibinfo {journal} {Phys. Rev. Lett.}}\
  }%
  \textbf{\bibinfo {volume} {89}},\ \bibinfo {pages} {248304} (\bibinfo {month}
  {Nov}\ \bibinfo {year} {2002})%
  \bibAnnoteFile{NoStop}{FuchsPRL2002}%
\bibitem{Felderhof83B}%
  \BibitemOpen
  \bibfield{author}{%
  \bibinfo {author} {\bibfnamefont{B.}~\bibnamefont{Felderhof}}\ and\ \bibinfo
  {author} {\bibfnamefont{R.}~\bibnamefont{Jones}},\ }%
  \bibfield{journal}{%
  \bibinfo {journal} {Faraday Discuss. Chem. Soc.}\ }%
  \textbf{\bibinfo {volume} {76}},\ \bibinfo {pages} {179} (\bibinfo {year}
  {1983})%
  \bibAnnoteFile{NoStop}{Felderhof83B}%
\bibitem{BraderPRL2007}%
  \BibitemOpen
  \bibfield{author}{%
  \bibinfo {author} {\bibfnamefont{J.~M.}\ \bibnamefont{Brader}}, \bibinfo
  {author} {\bibfnamefont{T.}~\bibnamefont{Voigtmann}}, \bibinfo {author}
  {\bibfnamefont{M.~E.}\ \bibnamefont{Cates}},\ and\ \bibinfo {author}
  {\bibfnamefont{M.}~\bibnamefont{Fuchs}},\ }%
  \bibfield{journal}{%
  \Doi{10.1103/PhysRevLett.98.058301}{\bibinfo {journal} {Phys. Rev. Lett.}}\
  }%
  \textbf{\bibinfo {volume} {98}},\ \bibinfo {pages} {058301} (\bibinfo {month}
  {Jan}\ \bibinfo {year} {2007})%
  \bibAnnoteFile{NoStop}{BraderPRL2007}%
\bibitem{BraderPRL2008}%
  \BibitemOpen
  \bibfield{author}{%
  \bibinfo {author} {\bibfnamefont{J.~M.}\ \bibnamefont{Brader}}, \bibinfo
  {author} {\bibfnamefont{M.~E.}\ \bibnamefont{Cates}},\ and\ \bibinfo {author}
  {\bibfnamefont{M.}~\bibnamefont{Fuchs}},\ }%
  \bibfield{journal}{%
  \Doi{10.1103/PhysRevLett.101.138301}{\bibinfo {journal} {Phys. Rev. Lett.}}\
  }%
  \textbf{\bibinfo {volume} {101}},\ \bibinfo {pages} {138301} (\bibinfo
  {month} {Sep}\ \bibinfo {year} {2008})%
  \bibAnnoteFile{NoStop}{BraderPRL2008}%
\bibitem{Henrich2009}%
  \BibitemOpen
  \bibfield{author}{%
  \bibinfo {author} {\bibfnamefont{O.}~\bibnamefont{Henrich}}, \bibinfo
  {author} {\bibfnamefont{F.}~\bibnamefont{Weysser}}, \bibinfo {author}
  {\bibfnamefont{M.~E.}\ \bibnamefont{Cates}},\ and\ \bibinfo {author}
  {\bibfnamefont{M.}~\bibnamefont{Fuchs}},\ }%
  \bibfield{journal}{%
  \Doi{10.1098/rsta.2009.0191}{\bibinfo {journal} {Physical and Engineering
  Sciences}}\ }%
  \textbf{\bibinfo {volume} {367}},\ \bibinfo {pages} {5033} (\bibinfo {month}
  {Dec.}\ \bibinfo {year} {2009})%
  \bibAnnoteFile{NoStop}{Henrich2009}%
\bibitem{Brady88}%
  \BibitemOpen
  \bibfield{author}{%
  \bibinfo {author} {\bibfnamefont{J.~F.}\ \bibnamefont{{Brady}}}\ and\
  \bibinfo {author} {\bibfnamefont{G.}~\bibnamefont{{Bossis}}},\ }%
  \bibfield{journal}{%
  \bibinfo {journal} {Annual Review of Fluid Mechanics}\ }%
  \textbf{\bibinfo {volume} {20}},\ \bibinfo {pages} {111} (\bibinfo {year}
  {1988})%
  \bibAnnoteFile{NoStop}{Brady88}%
\bibitem{Rjoy05}%
  \BibitemOpen
  \bibfield{author}{%
  \bibinfo {author} {\bibfnamefont{R.}~\bibnamefont{Adhikari}}, \bibinfo
  {author} {\bibfnamefont{K.}~\bibnamefont{Stratford}}, \bibinfo {author}
  {\bibfnamefont{M.~E.}\ \bibnamefont{Cates}},\ and\ \bibinfo {author}
  {\bibfnamefont{A.~J.}\ \bibnamefont{Wagner}},\ }%
  \bibfield{journal}{%
  \bibinfo {journal} {Europhysics Letters}\ }%
  \textbf{\bibinfo {volume} {71}},\ \bibinfo {pages} {473} (\bibinfo {year}
  {2005})%
  \bibAnnoteFile{NoStop}{Rjoy05}%
\bibitem{Groot97}%
  \BibitemOpen
  \bibfield{author}{%
  \bibinfo {author} {\bibfnamefont{R.~D.}\ \bibnamefont{Groot}}\ and\ \bibinfo
  {author} {\bibfnamefont{P.~B.}\ \bibnamefont{Warren}},\ }%
  \bibfield{journal}{%
  \bibinfo {journal} {Journal of Chemical Physics}\ }%
  \textbf{\bibinfo {volume} {107}},\ \bibinfo {pages} {4423} (\bibinfo {year}
  {1997})%
  \bibAnnoteFile{NoStop}{Groot97}%
\bibitem{Tanaka00}%
  \BibitemOpen
  \bibfield{author}{%
  \bibinfo {author} {\bibfnamefont{H.}~\bibnamefont{Tanaka}}\ and\ \bibinfo
  {author} {\bibfnamefont{T.}~\bibnamefont{Araki}},\ }%
  \bibfield{journal}{%
  \bibinfo {journal} {Phys.~Rev.\ Lett.}\ }%
  \textbf{\bibinfo {volume} {85}},\ \bibinfo {pages} {1338} (\bibinfo {year}
  {2000})%
  \bibAnnoteFile{NoStop}{Tanaka00}%
\bibitem{Ermak78}%
  \BibitemOpen
  \bibfield{author}{%
  \bibinfo {author} {\bibfnamefont{D.~L.}\ \bibnamefont{{Ermak}}}\ and\
  \bibinfo {author} {\bibfnamefont{J.~A.}\ \bibnamefont{{McCammon}}},\ }%
  \bibfield{journal}{%
  \bibinfo {journal} {Journal of Chemical Physics}\ }%
  \textbf{\bibinfo {volume} {69}},\ \bibinfo {pages} {1352} (\bibinfo {month}
  {Aug.}\ \bibinfo {year} {1978})%
  \bibAnnoteFile{NoStop}{Ermak78}%
\bibitem{Cichocki90sim}%
  \BibitemOpen
  \bibfield{author}{%
  \bibinfo {author} {\bibfnamefont{B.}~\bibnamefont{Cichocki}}\ and\ \bibinfo
  {author} {\bibfnamefont{H.}~\bibnamefont{K.}},\ }%
  \bibfield{journal}{%
  \bibinfo {journal} {Physica A}\ }%
  \textbf{\bibinfo {volume} {166}},\ \bibinfo {pages} {473} (\bibinfo {month}
  {Jul.}\ \bibinfo {year} {1990})%
  \bibAnnoteFile{NoStop}{Cichocki90sim}%
\bibitem{Heyes93}%
  \BibitemOpen
  \bibfield{author}{%
  \bibinfo {author} {\bibfnamefont{D.~M.}\ \bibnamefont{Heyes}}\ and\ \bibinfo
  {author} {\bibfnamefont{J.~R.}\ \bibnamefont{Melrose}},\ }%
  \bibfield{journal}{%
  \bibinfo {journal} {Journal of Non-Newtonian Fluid Mechanics}\ }%
  \textbf{\bibinfo {volume} {46}},\ \bibinfo {pages} {1} (\bibinfo {month}
  {Jan.}\ \bibinfo {year} {1993})%
  \bibAnnoteFile{NoStop}{Heyes93}%
\bibitem{Sillescu94}%
  \BibitemOpen
  \bibfield{author}{%
  \bibinfo {author} {\bibfnamefont{W.}~\bibnamefont{Schaertl}}\ and\ \bibinfo
  {author} {\bibfnamefont{H.}~\bibnamefont{Sillescu}},\ }%
  \bibfield{journal}{%
  \bibinfo {journal} {Journal of Statistical Physics}\ }%
  \textbf{\bibinfo {volume} {74}},\ \bibinfo {pages} {687} (\bibinfo {month}
  {Feb.}\ \bibinfo {year} {1994})%
  \bibAnnoteFile{NoStop}{Sillescu94}%
\bibitem{Foss2000}%
  \BibitemOpen
  \bibfield{author}{%
  \bibinfo {author} {\bibfnamefont{D.~R.}\ \bibnamefont{{Foss}}}\ and\ \bibinfo
  {author} {\bibfnamefont{J.~F.}\ \bibnamefont{{Brady}}},\ }%
  \bibfield{journal}{%
  \bibinfo {journal} {Journal of Fluid Mechanics}\ }%
  \textbf{\bibinfo {volume} {407}},\ \bibinfo {pages} {167} (\bibinfo {month}
  {Mar.}\ \bibinfo {year} {2000})%
  \bibAnnoteFile{NoStop}{Foss2000}%
\bibitem{Strating99}%
  \BibitemOpen
  \bibfield{author}{%
  \bibinfo {author} {\bibfnamefont{P.}~\bibnamefont{Strating}},\ }%
  \bibfield{journal}{%
  \bibinfo {journal} {Physical Review E}\ }%
  \textbf{\bibinfo {volume} {59}},\ \bibinfo {pages} {2175} (\bibinfo {month}
  {Feb.}\ \bibinfo {year} {1999})%
  \bibAnnoteFile{NoStop}{Strating99}%
\bibitem{Tao06}%
  \BibitemOpen
  \bibfield{author}{%
  \bibinfo {author} {\bibfnamefont{Y.-G.}\ \bibnamefont{Tao}}, \bibinfo
  {author} {\bibfnamefont{W.~K.}\ \bibnamefont{den Otter}}, \bibinfo {author}
  {\bibfnamefont{J.~K.~G.}\ \bibnamefont{Dhont}},\ and\ \bibinfo {author}
  {\bibfnamefont{W.~J.}\ \bibnamefont{Briels}},\ }%
  \bibfield{journal}{%
  \bibinfo {journal} {Journal of Chemical Physics}\ }%
  \textbf{\bibinfo {volume} {124}},\ \bibinfo {pages} {134906} (\bibinfo {year}
  {2006})%
  \bibAnnoteFile{NoStop}{Tao06}%
\bibitem{BD4HS}%
  \BibitemOpen
  \bibfield{author}{%
  \bibinfo {author} {\bibfnamefont{A.}~\bibnamefont{{Scala}}}, \bibinfo
  {author} {\bibfnamefont{C.}~\bibnamefont{{De~Michele}}},\ and\ \bibinfo
  {author} {\bibfnamefont{T.}~\bibnamefont{{Voigtmann}}},\ }%
  \bibfield{journal}{%
  \bibinfo {journal} {Journal of Chemical Physics}\ }%
  \textbf{\bibinfo {volume} {126}},\ \bibinfo {pages} {134109} (\bibinfo
  {month} {Apr.}\ \bibinfo {year} {2007})%
  \bibAnnoteFile{NoStop}{BD4HS}%
\bibitem{KannanBook}%
  \BibitemOpen
  \bibfield{author}{%
  \bibinfo {author} {\bibfnamefont{D.}~\bibnamefont{Kannan}}\ and\ \bibinfo
  {author} {\bibfnamefont{V.}~\bibnamefont{Lakshmikantham}},\ }%
  \emph{\bibinfo {title} {Handbook of stochastic analysis and applications}}\
  (\bibinfo {publisher} {Marcel Dekker},\ \bibinfo {year} {2002})\
  Chap.~\bibinfo {chapter} {5}%
  \bibAnnoteFile{NoStop}{KannanBook}%
\bibitem{KloedenPlatenBook}%
  \BibitemOpen
  \bibfield{author}{%
  \bibinfo {author} {\bibfnamefont{P.~E.}\ \bibnamefont{Kloeden}}\ and\
  \bibinfo {author} {\bibfnamefont{E.}~\bibnamefont{Platen}},\ }%
  \emph{\bibinfo {title} {Numerical solution of stochastic differential
  equations}},\ \bibinfo {edition} {3rd}\ ed.,\ \bibinfo {series} {Applications
  of Mathematics}, Vol.~\bibinfo {volume} {23}\ (\bibinfo {publisher}
  {Springer},\ \bibinfo {year} {1999})%
  \bibAnnoteFile{NoStop}{KloedenPlatenBook}%
\bibitem{Kramers1940}%
  \BibitemOpen
  \bibfield{author}{%
  \bibinfo {author} {\bibfnamefont{H.}~\bibnamefont{Kramers}},\ }%
  \bibfield{journal}{%
  \Doi{10.1016/S0031-8914(40)90098-2}{\bibinfo {journal} {Physica}}\ }%
  \textbf{\bibinfo {volume} {7}},\ \bibinfo {pages} {284 } (\bibinfo {year}
  {1940}),\ ISSN \bibinfo {issn} {0031-8914}%
  \bibAnnoteFile{NoStop}{Kramers1940}%
\bibitem{Forbert00}%
  \BibitemOpen
  \bibfield{author}{%
  \bibinfo {author} {\bibfnamefont{H.~A.}\ \bibnamefont{Forbert}}\ and\
  \bibinfo {author} {\bibfnamefont{S.~A.}\ \bibnamefont{Chin}},\ }%
  \bibfield{journal}{%
  \bibinfo {journal} {Physical Review E}\ }%
  \textbf{\bibinfo {volume} {63}},\ \bibinfo {pages} {016703} (\bibinfo {month}
  {Dec.}\ \bibinfo {year} {2000})%
  \bibAnnoteFile{NoStop}{Forbert00}%
\bibitem{Chin05}%
  \BibitemOpen
  \bibfield{author}{%
  \bibinfo {author} {\bibfnamefont{S.~A.}\ \bibnamefont{{Chin}}},\ }%
  \bibfield{journal}{%
  \bibinfo {journal} {Physical Review E}\ }%
  \textbf{\bibinfo {volume} {71}},\ \bibinfo {pages} {016703} (\bibinfo {month}
  {Jan.}\ \bibinfo {year} {2005})%
  \bibAnnoteFile{NoStop}{Chin05}%
\bibitem{RiskenBook}%
  \BibitemOpen
  \bibfield{author}{%
  \bibinfo {author} {\bibfnamefont{H.}~\bibnamefont{{Risken}}},\ }%
  \emph{\bibinfo {title} {{The Fokker-Planck equation. Methods of solution and
  applications}}}\ (\bibinfo {publisher} {Springer Series in Synergetics,
  Berlin, New York: Springer, |c1989, 2nd ed.},\ \bibinfo {year} {1989})%
  \bibAnnoteFile{NoStop}{RiskenBook}%
\bibitem{RapaBook}%
  \BibitemOpen
  \bibfield{author}{%
  \bibinfo {author} {\bibfnamefont{D.~C.}\ \bibnamefont{{Rapaport}}},\ }%
  \emph{\bibinfo {title} {{The Art of Molecular Dynamics Simulation}}}\
  (\bibinfo {publisher} {Cambridge University Press},\ \bibinfo {year} {2004})%
  \bibAnnoteFile{NoStop}{RapaBook}%
\bibitem{Kennedy99}%
  \BibitemOpen
  \bibfield{author}{%
  \bibinfo {author} {\bibfnamefont{A.~D.}\ \bibnamefont{Kennedy}},\ }%
  \bibfield{journal}{%
  \bibinfo {journal} {Parallel Computing}\ }%
  \textbf{\bibinfo {volume} {7}},\ \bibinfo {pages} {284} (\bibinfo {month}
  {Apr.}\ \bibinfo {year} {1999})%
  \bibAnnoteFile{NoStop}{Kennedy99}%
\bibitem{ATBook}%
  \BibitemOpen
  \bibfield{author}{%
  \bibinfo {author} {\bibfnamefont{M.~P.}\ \bibnamefont{Allen}}\ and\ \bibinfo
  {author} {\bibfnamefont{D.~J.}\ \bibnamefont{Tildesley}},\ }%
  \emph{\bibinfo {title} {Computer Simulation of Liquids}},\ \bibinfo {edition}
  {2nd}\ ed.\ (\bibinfo {publisher} {Clarendon Press, Oxford},\ \bibinfo {year}
  {1987})%
  \bibAnnoteFile{NoStop}{ATBook}%
\bibitem{TrotterPAMS59}%
  \BibitemOpen
  \bibfield{author}{%
  \bibinfo {author} {\bibfnamefont{H.~F.}\ \bibnamefont{Trotter}},\ }%
  \bibfield{journal}{%
  \bibinfo {journal} {Proc. Amer. Math. Soc.}\ }%
  \textbf{\bibinfo {volume} {10}},\ \bibinfo {pages} {545} (\bibinfo {year}
  {1959})%
  \bibAnnoteFile{NoStop}{TrotterPAMS59}%
\bibitem{DonevSIM2009}%
  \BibitemOpen
  \bibfield{author}{%
  \bibinfo {author} {\bibfnamefont{A.}~\bibnamefont{Donev}},\ }%
  \bibfield{journal}{%
  \bibinfo {journal} {SIMULATION}\ }%
  \textbf{\bibinfo {volume} {85}},\ \bibinfo {pages} {229} (\bibinfo {year}
  {2009})%
  \bibAnnoteFile{NoStop}{DonevSIM2009}%
\bibitem{HardBrown}%
  \BibitemOpen
  \bibfield{author}{%
  \bibinfo {author} {\bibfnamefont{A.}~\bibnamefont{Scala}},\ }%
  \enquote{\bibinfo {title} {Simulation of hard brownian and granular
  particles},}\  (\bibinfo {year} {2008}),\
  \url{http://gna.org/projects/hardbrown}%
  \bibAnnoteFile{NoStop}{HardBrown}%
\end{thebibliography}%

\newpage

\end{document}